\documentclass[lettersize,journal]{IEEEtran}
\usepackage{amsmath,amsfonts}
\usepackage{algorithmic}
\usepackage{algorithm}
\usepackage{float}
\usepackage{url}
\usepackage{graphicx}
\usepackage{cite}
\usepackage[dvipsnames]{xcolor}
\usepackage{multirow}
\usepackage{tabularray}
\usepackage{colortbl}
\usepackage{hhline}
\usepackage{latexsym}
\usepackage{bbding}
\usepackage{pifont}
\usepackage{subfigure}
\usepackage{adjustbox}
\usepackage{flushend}

\usepackage[mathscr]{urwchancal}

\makeatletter
\newcommand{\specialcell}[1]{\ifmeasuring@#1\else\omit$\displaystyle#1$\ignorespaces\fi}

\makeatother
\definecolor{lightred}{RGB}{255,220,220}
\definecolor{lightgreen}{RGB}{255,255,255}
\definecolor{lightgray}{RGB}{225,225,225}
\newcolumntype{P}[1]{>{\centering\arraybackslash}p{#1}}
\newcolumntype{M}[1]{>{\centering\arraybackslash}m{#1}}
\begin{document}

\title{Flip-KLJN: Random Resistance Flipping for Noise-Driven Secure Communication}

\author{Recep~A.~Tasci,~\IEEEmembership{Graduate~Student~Member,~IEEE,}
        Ibrahim~Yildirim,~\IEEEmembership{Member,~IEEE,}
        Ertugrul~Basar,~\IEEEmembership{Fellow,~IEEE}

\thanks{R. A. Tasci is with the Communications Research and Innovation Laboratory (CoreLab), Department of Electrical and Electronics Engineering, Koc University, Sariyer 34450, Istanbul, Turkey. (e-mail: rtasci20@ku.edu.tr).}
\thanks{I. Yildirim is with the Faculty of Electrical and Electronics Engineering, Istanbul Technical University, Istanbul, Turkey, and also with the Department of Electrical and Computer Engineering, McGill University, Montreal, QC, Canada. (e-mail: yildirimib@itu.edu.tr)}
\thanks{E. Basar is with the Department of Electrical Engineering, Tampere University, 33720 Tampere, Finland, on leave from the Department of Electrical and Electronics Engineering, Koc University, 34450 Sariyer, Istanbul, Turkey (e-mails: ertugrul.basar@tuni.fi, ebasar@ku.edu.tr)}
}
\markboth{ }%
{Tasci \MakeLowercase{\textit{et al.}}: Flip-KLJN: Random Resistance Flipping for Noise-Driven Secure Communication}


\maketitle

\begin{abstract}
The information-theoretically (unconditionally) secure Kirchhoff-law-Johnson-noise (KLJN) bit exchange protocol uses two identical resistor pairs with high ($H$) and low ($L$) resistance values, driven by Gaussian noise generators emulating Johnson noise with a high common temperature. The resulting mean-square noise voltage on the wire connecting Alice and Bob has three levels: low ($L/L$), intermediate ($H/L$ or $L/H$), and high ($H/H$), and secure key sharing is achieved at the intermediate level ($L/H$ or $H/L$). This paper introduces the \textit{Flip-KLJN} scheme, where a pre-agreed intermediate level, such as $H/L$, triggers a flip of the bit map value during the bit exchange period. For Eve, the bit map flips appear random. Thus, the formerly discarded $H/H$ and $L/L$ situations can also have a pre-agreed bit value mapping, which flips together with the original bit mapping. Thus, Flip-KLJN doubles the key rate and ensures that all three levels on the wire are indistinguishable for Eve. Bit error probabilities are addressed through analytic calculations and computer simulations.
\end{abstract}


\section{Introduction}
\IEEEPARstart{T}{he} advent of next-generation communication systems represents a critical shift in the landscape of wireless communications. This shift is necessitated by emerging demands for heightened security \cite{9419781,CIT-036}, enhanced energy efficiency \cite{9758764}, improved spectral efficiency and latency, and increased stealth in various scenarios. One of the key 6G technologies relevant to these needs is encrypted communication, ensuring that sensitive information remains confidential and protected from unauthorized access \cite{7393435}. Effective key generation methods provide robust protection against eavesdropping and cyberattacks. The importance of secure key generation continues to grow with the expanding complexity of communication networks. 

Recent research underscores RSA encryption's vulnerability to quantum computers. A Chinese team recently used a quantum computer to break RSA-$2048$ for $22$-bit keys, showing that future advances in quantum hardware could threaten larger keys \cite{CSO2024}. This highlights the need for quantum-resistant cryptography and secure alternatives like Quantum Key Distribution (QKD), which uses quantum mechanics to provide secure keys. However, QKD’s high cost and scalability issues limit its adoption. In contrast, KLJN schemes are low-cost and robust for short-range applications\cite{kmj}, and can be demonstrated for various use cases\cite{7527437}.

In this context, key generation with thermal noise emerges as a compelling alternative to current technologies, effectively addressing security and vulnerability requirements. Thermal noise, an inherent characteristic of electronic circuits, significantly impacts the performance of communication systems. It arises from the random vibration of charge carriers within circuit conductors, and its intensity is directly proportional to the device's temperature, resistance, and bandwidth \cite{100096109}. It is a ubiquitous phenomenon in communication systems, affecting the performance of data transmission and reception, such as the bit error rate (BER) and signal-to-noise ratio (SNR). To mitigate the impact of thermal noise in communication systems, various techniques are employed, including increasing transmission power, utilizing low-noise amplifiers, narrowing system bandwidth, implementing robust modulation schemes \cite{7509396,8004416}, applying error-correcting codes, deploying high-gain directional antennas, and maintaining precise temperature control.

Unlike conventional methods, \cite{kish2005stealth} leverages inherent background noise, which traditionally impedes communication, to its advantage. This innovative method, which uses two different resistances to create distinct noise spectra, is particularly advantageous in scenarios where conventional transmission power is limited or the security and undetectability of communication are paramount. Advancing this concept further, Kirchhoff-Law-Johnson-(like)-Noise (KLJN) secure key exchange scheme using the thermal noises of two pairs of resistors is proposed in \cite{kish2006totally,kish2006totally2} to achieve unconditionally secure communication by utilizing the Kirchhoff’s law. The KLJN loop requires Gaussian noise sources; thus, non-Gaussian noise types does not establish the loop \cite{100096109}. In \cite{mingesz2008johnson}, the authors focused on the practical applications of KLJN communicators at several distances and data rates, and highlighted that the information leakage in the KLJN communicators is comparatively less significant than in quantum communication systems. Furthermore, it has been established that the use of Johnson-like noise, whether naturally generated or externally produced, is essential for secure key exchange in KLJN systems \cite{KISH20102140}. The study of \cite{kish2013enhanced} introduced seven new variants of the KLJN secure key exchange scheme, which provide enhanced security in non-ideal conditions. The study of \cite{9980386} also examined the KLJN secure bit exchange scheme by proposing two new detectors using voltage and current measurements to reduce BER. A closed-form expression for the bit error probability (BEP) in a thermal noise modulation-based wireless communication system using a maximum likelihood (ML) detector is derived in \cite{10706852}, along with the calculation of the detector's optimal threshold value. Moreover, taking inspirations from KLJN designs, the study of \cite{10373568} proposed alternative noise modulation designs, exploiting non-coherent detection and time diversity by using noise variance for digital communication.

The KLJN design uses thermal noise generated by two resistors, one from Alice and the other from Bob. Each party, Alice and Bob, possesses two resistors: a low resistance ``$L$" and a high resistance ``$H$". They randomly select one resistor to connect to a common wire for secure key exchange \cite{kish2006totally,kish2006totally2}. This configuration produces three possible noise levels on the line: low ($L/L$), high ($H/H$), and intermediate ($H/L$ or $L/H$). For instance, $L/H$ indicates that Alice uses a low resistor while Bob uses a high resistor. Although Eve can detect this intermediate noise level, the specific resistances remain undetectable, ensuring unconditional security. However, Eve can decipher the signal if the noise level is either low or high, rendering these noise levels insecure. Alice and Bob estimate the noise power on the line to determine each other's resistor and generate the key bit $0$ if the resistors are $L/H$, and the key bit $1$ if the resistors are $H/L$, rejecting the $L/L$ and $H/H$ cases to avoid generating insecure key bits. As a result, half of the bits are discarded, as they do not contribute to key generation in the $L/L$ and $H/H$ cases, but only in the $L/H$ and $H/L$ cases. To address this, the ``Flip-KLJN" design is proposed to ensure unconditional security regardless of the noise level. This new design shares some similarities with the enhanced KLJN variant presented in \cite{kish2013enhanced}, as both aim to further confuse Eve by manipulating bit mappings. However, in Flip-KLJN, bit interpretations are randomly updated during one of the intermediate noise level case, eliminating the need for a prior key. This approach ensures that even $L/L$ and $H/H$ cases can confuse Eve, achieving unconditional security for all bit combinations while doubling the key rate.

In light of these, the primary contributions of this article can be outlined as follows:
\begin{itemize}
    \item We propose a novel design called Flip-KLJN. In this design, the resistor pair mapping is flipped systematically based on the generated bits, and this reversal is done under specific conditions that a potential eavesdropper cannot detect. This approach confuses eavesdroppers and ensures unconditional security comparable to quantum secrecy for all of the noise levels.
    \item The proposed design doubles the key rate of the KLJN secure bit exchange scheme by purposefully injecting a controlled element of unpredictability into the communication process. This enhancement stems from the inherent feature of Flip-KLJN, where all transmitted bits achieve a state of absolute security and remain undetectable by potential eavesdroppers. In contrast, classical KLJN only ensures the security of $50\%$ of transmitted bits, emphasizing the innovative and effective role of the Flip-KLJN algorithm in key expansion. Moreover, the proposed design neither enhances nor reduces security, maintaining the same security level as the classical KLJN design.
    \item The proposed design can also be utilized for two-way secure data transfer. In this configuration, two entities can simultaneously communicate by both receiving and transmitting information bits. However, to ensure security, the bits must be scrambled using a randomization algorithm. This approach makes the eavesdropper (Eve) unable to decode the transmitted bits by observing patterns in noise levels, even if she knows the randomization algorithm, as she will still be unable to correctly interpret the data.
    \item The proposed design is integrated into various detectors in order to reduce the BER difference between the Flip-KLJN and KLJN designs. This strategic integration effectively lessens the chance of error propagation, leading to a more balanced and improved BER performance.
    \item In addition to elaborative and detailed theoretical BER calculations of the proposed design, we provide a valuable contribution by conducting comprehensive computer simulations. This approach not only strengthens the credibility of the proposed design but also facilitates a more nuanced assessment of its performance under varying conditions.
\end{itemize}

This article is structured as follows. Section II provides a brief overview of the variants of KLJN secure bit exchange design and enhanced KLJN detectors. Section III explains the Flip-KLJN design in more detail. Section IV presents the system model of the proposed design and theoretical BER calculations. Section V contains our numerical results, and Section VI concludes the paper.

\section{Enhanced Variants and Detectors of Classical KLJN Secure Bit Exchange Scheme}
In this section, we provide a brief overview some of the enhanced variants of classical KLJN secure bit exchange scheme and some detectors using joint voltage and current measurements.
\vspace{-5pt}
\subsection{Enhanced Variants of KLJN}

\subsubsection{Intelligent KLJN (iKLJN) Scheme}
The iKLJN system allows Alice and Bob to use a shorter KLJN clock period by utilizing their knowledge of their own resistor values and the stochastic time functions of their own noise. By subtracting their own noise contributions, they generate reduced channel noise that does not contain their own noise components, helping them to better determine the correct resistance value at the other end. This approach reduces Eve's ability to gather statistics within the limited time window.

\subsubsection{Multiple KLJN (MKLJN) Scheme}
In the MKLJN system, Alice and Bob have publicly known identical sets of different resistors that are randomly chosen and connected to the line. The bit interpretation of the different resistor combinations is defined in a publicly known truth table. This scheme requires Eve to identify the actual resistor values at both ends accurately, which is more challenging and results in enhanced security.

\subsubsection{Keyed KLJN (KKLJN) Scheme}\label{subsection:KKLJN}
The KKLJN scheme shares a time-dependent truth table for bit interpretation using a previously shared secure key. This method ensures that even if Eve guesses the current key, the security of subsequent keys remains high. Eve must know the former key to understand the bit interpretations of the resistor situations, making it significantly harder for her to compromise the key exchange process, which progressively reduces her information about the new key\cite{kish2013enhanced}.

\subsection{Enhanced KLJN Detectors}
Due to the Second Law of Thermodynamics and the Gaussian nature of the noises, where the cross-correlation between two Gaussian processes with zero mean is zero, resulting in statistical independence, the current on the wire can be utilized as supplementary information for bit detection while the voltage measurement is also being used.

Leveraging this, a joint voltage and current measurement-based detector is proposed in \cite{4578679}. This approach selects the cumulative measurement output with the smallest associated error, utilizing the fact that the voltage-based method exhibits minimum error probability in scenarios where the current-based method has maximum error probability, and vice versa.

In \cite{9980386}, the author thoroughly examines the theoretical BER calculations for the KLJN secure bit exchange scheme proposed in \cite{kish2006totally,kish2006totally2}. The author also introduces two new KLJN detectors that utilize samples from both voltage and current noise waveforms. One of these detectors smartly raises an error flag when the decisions about voltage and current bits do not match. It is important to note that this detector is similar to the one proposed in \cite{4578679}, and we call this detector as joint voltage-current detector (JVCD) in this paper. Moving to the other detector, it goes a step further by recognizing the prevalent error events for voltage and current measurements. Consequently, this detector enables Alice and Bob to select the types of measurements they make based on their individual resistors. Specifically, if Alice's (or Bob's) resistor is $L$, the detector prioritizes current measurements. Conversely, if the resistor is $H$, voltage measurements take precedence. This decision is based on the recognition that the occurrence likelihood of $L/L\leftrightarrow L/H$ (Alice's/Bob's resistor) and $H/L\leftrightarrow H/H$ error events is lower for current and voltage measurements, respectively. These detectors significantly enhance the robustness of communication, leading to a considerable drop in the BER.
\begin{figure*}[t!]
    \centering
    \includegraphics[width=0.98\textwidth]{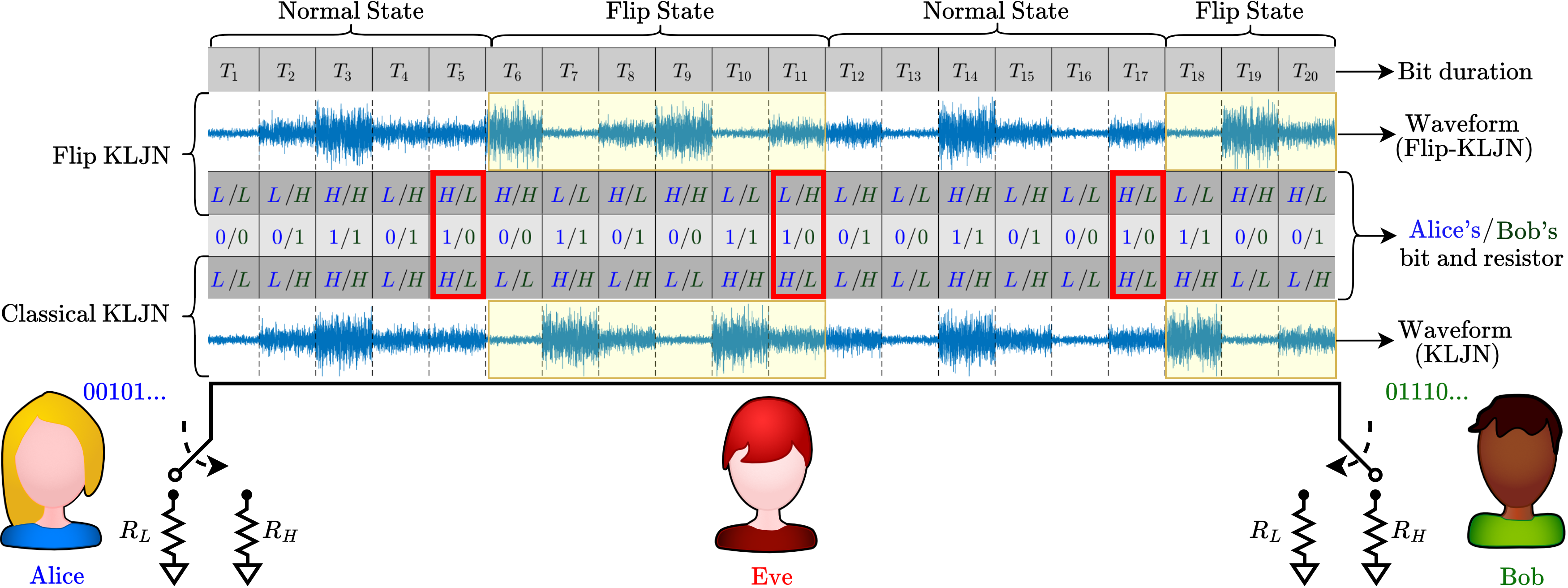}\vspace{-5pt}
    \caption{Representation of the noise levels in Normal State and Flip State of Flip-KLJN scheme and transition of states.} 
    \label{fig:systemmdl}
\end{figure*}
\section{FLIP-KLJN Scheme}
Similar to the classical KLJN, our proposed design involves two communicating entities, namely Alice and Bob. They select the resistors randomly and connect them to a wire channel during each transmission interval with the help of a switch. If both Alice and Bob possess the resistor $L$, the noise level in the wire will be relatively low. Conversely, if they both possess the resistor $H$, the noise level in the wire will be comparatively high. The security of the bit exchange process is based on the distinguishability of the resistors at both ends; when they are different, an intermediate mean-square noise voltage level appears on the line. Despite the potential detection of this intermediate noise level by an eavesdropper (Eve), the resistors of Alice and Bob remain undetectable. This lack of comprehension ensures an unconditional security level comparable to quantum secrecy. Conversely, Eve can decipher the signal's content if the noise level on the wire is non-intermediate, i.e., when both Alice and Bob selects $L$ or $H$, resulting in low or high noise levels, respectively. Thus, the traditional KLJN scheme discards these insecure bits, leading to a significantly reduced key rate. To address this limitation and the inherent vulnerability associated with high and low noise levels, we propose a novel approach called Flip-KLJN, which ensures unconditional security for key generation even when the noise level is non-intermediate. It is important to note that Flip-KLJN is not more secure than the KLJN scheme, as KLJN already discards non-secure bits. Instead, Flip-KLJN extends the key length by using the discarded bits in KLJN and remains equally resilient to the potential attacks \cite{ferdous2023transient,melhem2019static,kish2014elimination}.

To clarify the explanations and the theoretical calculations for this system, we map the resistors used by Alice and Bob to specific bit values. It is important to note that these bit values are not information bits; they remain random for key generation. This mapping is solely for associating the resistors with corresponding bit values. Additionally, after the key transmission is completed, Alice and Bob can use Alice's random bits to establish a fully secure common key. The mapping is illustrated in Table \ref{tablebitsress} for clarity.
\begin{table}[t!]
\centering
\caption{\label{tablebitsress} Noise Power and Eve's Decision Based on the Resistor-Bit Mapping}
\begin{tblr}{
  cell{2}{1} = {r=4}{},
  cell{6}{1} = {r=4}{},
  vline{1,2,3,4,5,6}={1.1pt},
  hline{1,2,6,10} = {1.1pt},
  hline{3-9} = {2-5}{},
  colspec = {X[c,m]X[c,m]X[c,m]X[c,m]},
  stretch = 0.1,
  rowsep = 1pt,
}
\SetCell{bg=lightgray}{\textbf{State}} & \SetCell{bg=lightgray}{\textbf{Bits}\\\textbf{(Alice/Bob)}} & \SetCell{bg=lightgray}{\textbf{Resistor}\\\textbf{Mapping}\\\textbf{(Alice/Bob)}} & \SetCell{bg=lightgray}{\textbf{Noise Power}}\\
\SetCell{bg=lightgray}{ \textbf{Normal}\\\textbf{State}} & $0/0$ & $L/L$ & {Low}\\
 & $0/1$ & $L/H$ & {Medium}\\
 & $1/0$ & $H/L$ & {Medium}\\
 & $1/1$ & $H/H$ & {High}\\
\SetCell{bg=lightgray}{\textbf{Flip}\\\textbf{State}} & $0/0$ & $H/H$ & {High}\\
 & $0/1$ & $H/L$ & Medium\\
 & $1/0$ & $L/H$ & {Medium}\\
 & $1/1$ & $L/L$ & {Low}
\end{tblr}\vspace{-10pt}
\end{table}

As depicted in Table \ref{tablebitsress}, there are two states: the ``Normal" state and the ``Flip" state. In the Normal state, Alice and Bob map the resistors as follows: $L$ to bit $0$ and $H$ to bit $1$. In the Flip state, the mapping is reversed: $H$ to bit $0$ and $L$ to bit $1$. They switch between these states simultaneously when Alice's and Bob's random bits are $1/0$, respectively. Regardless of whether they are in the Flip or Normal state, the noise level on the wire remains intermediate. Specifically, in the Normal state, $1/0$ corresponds to $H/L$, and in the Flip state, $1/0$ corresponds to $L/H$. This ensures that Eve cannot detect the state switching, as it occurs at the intermediate noise level. Consequently, the low and high noise power cases are fully secure, as Eve cannot determine whether Alice and Bob are using $L/L$ or $H/H$ since she is unaware of their current state. It is important to remember that in the classical KLJN scheme, Eve does not attempt to estimate the bits when the noise power is low or high, as these insecure bits are discarded by Alice and Bob. Given that Eve is aware of the proposed scheme, she may attempt to estimate the bits even when the noise power on the wire is low or high, knowing that key bits can still be generated in this case. 

The switching process is illustrated in Fig. \ref{fig:systemmdl}, where Alice's random bits are depicted in blue, while Bob's random bits are shown in green. If $1/0$ is observed in the current bit duration, the state switching operation takes place in the subsequent bit duration. As an illustration, when the random bits are $0/0$ and $1/1$, the noise levels are low and high, respectively, in the Normal state. However, in the Flip state, this arrangement is reversed, with noise levels being high for $0/0$ and low for $1/1$. The noise level consistently remains at an intermediate level when transmitting the random bits $0/1$ or $1/0$. This is because, in both cases, resistors $L$ and $H$ are utilized individually at the terminals, maintaining an intermediate noise level. After a while, when the random bits are $1/0$, Alice and Bob return back to their previous state, where they stay at this state until the bits become $1/0$ again.

The proposed design extends the key length by a factor of two compared to the classical KLJN. On the other hand, the Flip-KLJN design exhibits a slightly inferior BER performance compared to the KLJN detector in \cite{9980386}. This discrepancy arises from the probability of erroneous detection on one of the terminals, resulting in failure to execute state switching when necessary, thereby leading to error propagation due to state mismatch. Fortunately, the natural resilience of the proposed design corrects error propagation approximately every two bits, providing a very strong framework. Furthermore, compared to the classical KLJN design, the proposed design does not require a significantly more complex circuit. It only necessitates a basic, low-cost decision circuit to determine whether the flipping operation occurs while all other components, as well as the synchronization process, remain identical to those in the classical KLJN system.
\section{System Model and Performance Analysis}

Our proposed design entails two communicating terminals connected with a wire, namely Alice and Bob, each having two resistors. Initially, they select a low-valued resistor, $L$, or a high-valued resistor, $H$, based on the random bits ($b_A$ and $b_B$ for Alice and Bob, respectively) and subsequently connect them to a wire channel during each transmission interval with the assistance of a switch. The connection of two resistors causes three possible thermal noise level on the wire, which are modeled as white Gaussian noise process with a mean of $0$ and variance of $\sigma^2\in\{\sigma_{00}^2,\sigma_{01}^2,\sigma_{11}^2\}$ where $\sigma_{00}^2$, $\sigma_{01}^2$, and $\sigma_{11}^2$ holds for the noise power on the wire when Alice$/$Bob uses $L/L$, $L/H$ or $H/L$, and $H/H$, respectively, and defined as follows\cite{9980386}:
\begin{align}\label{Eq:1}
    \sigma_{00}^2=4kT\dfrac{R_LR_L}{R_L+R_L}\Delta f,\nonumber\\
    \sigma_{01}^2=4kT\dfrac{R_LR_H}{R_L+R_H}\Delta f,\nonumber\\
    \sigma_{11}^2=4kT\dfrac{R_HR_H}{R_H+R_H}\Delta f,
\end{align}
where $k$ is the Boltzmann’s constant, which is $1.38\times 10^{-23}$ joules per Kelvin, $T$ is the temperature in Kelvin, $\Delta f$ is the bandwidth in Hz, and $R_L$ and $R_H$ are the resistance values for $L$ and $H$, respectively.
\begin{figure*}[t!]
    \centering
    \includegraphics[width=\textwidth]{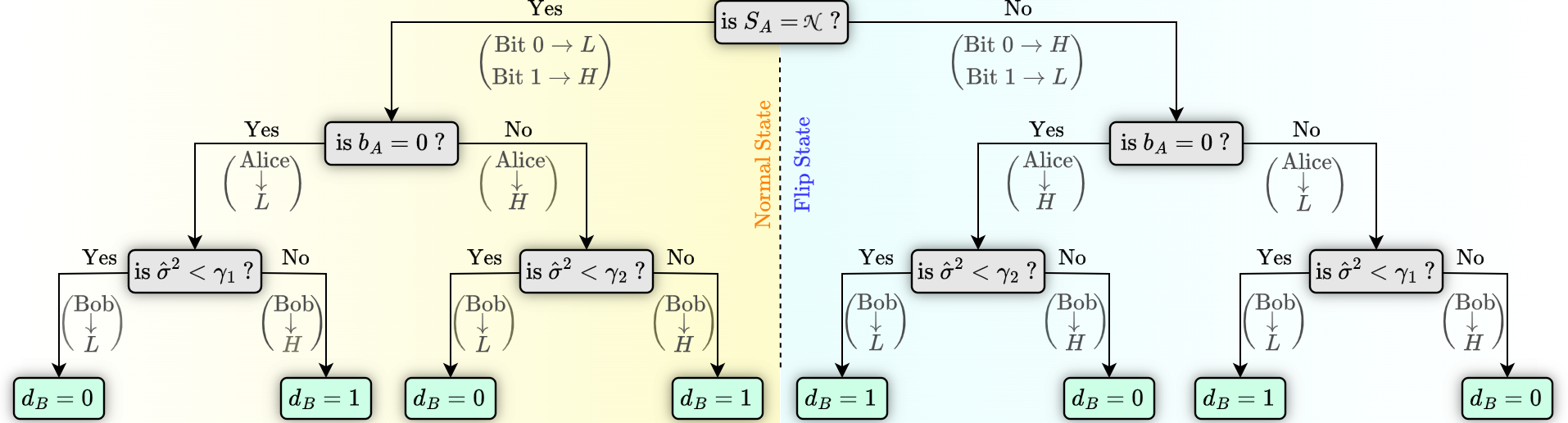}
    \caption{Decision tree for Alice's detection mechanism.} 
    \label{fig:dectree}
\end{figure*}

We now define $\hat{\sigma}^2$, which is the estimated voltage variance by Alice and Bob. It is calculated by taking $N$ samples from the wire and expressed as
\begin{align}
    \hat{\sigma}^2=\dfrac{1}{N}\sum_{k=1}^N x_k^2,
\end{align}
where $x_k\sim \mathcal{N}(0,\sigma^2)$ represents the $k$th independent noise sample. One can easily observe that $\mathbb{E}[\hat{\sigma}^2]=\sigma^2$, where $\mathbb{E}[\cdot]$ stands for the expectation. 
In an ideal scenario, $\hat{\sigma}^2$ follows a chi-square distribution. However, the central limit theorem (CLT) ensures that for sufficiently large $N$, $\hat{\sigma}^2$ approximates a Gaussian distribution as $\hat{\sigma}^2\sim \mathcal{N}(\sigma^2,2\sigma^4/N)$. Due to its high skewness, the chi-square distribution converges to a Gaussian more slowly than symmetric distributions \cite{box1978statistics}. Nonetheless, for the proposed design, $N > 50$ may provide a reasonable approximation. We establish two thresholds to determine whether $\hat{\sigma}^2$ lies closer to the three specified noise variances in (\ref{Eq:1}), defined as $\gamma_1=\beta\sigma_{00}^2$ and $\gamma_2=\kappa\sigma_{00}^2$. Here, $\beta$ and $\kappa$ represent certain constants, determined numerically to optimize the BER, as deriving an analytical solution is not the focus of this paper ($\eta$ and $\xi$ are the JVCD counterparts \cite{9980386}). It is worth noting that $\sigma_{00}^2<\gamma_1<\sigma_{01}^2<\gamma_2<\sigma_{11}^2$. Here, we define $\alpha=R_H/R_L$, and it follows that $\sigma_{01}^2=(2\alpha/(1+\alpha))\sigma_{00}^2$, $\sigma_{11}^2=\alpha\sigma_{00}^2$, and $1<\beta<\frac{2\alpha}{1+\alpha}<\kappa<\alpha$.

Each selection of resistor pairs results in different noise variances on the wire, consequently affecting $\hat{\sigma}^2$ as well. Decisions are based on $\hat{\sigma}^2$, $\gamma_1$, and $\gamma_2$. However, the most influential factor is the states of Alice and Bob in the Flip-KLJN design, as the mapping of resistances to bits changes frequently. Initially, bit $0$ and bit $1$ are represented by selecting resistances $L$ and $R$, respectively, defining the state known as the Normal State ($\mathscr{N}$). After several bit durations, Alice and Bob apply a reversal mapping to their resistances, as shown in Table \ref{tablebitsress}. This reversal is triggered by the transmission of a predefined random bit pair, $b_A/b_B=1/0$, which switches the system to the Flip State ($\mathscr{F}$), where $H$ now corresponds to bit $0$ and $L$ to bit $1$. Conversely, the state transitions back to $\mathscr{N}$ if the prior state was $\mathscr{F}$. Both Alice's and Bob's states ($S_A$ and $S_B$) are flipped for subsequent bits after $b_A/b_B=1/0$. When the flipping operation occurs, $\sigma^2$ may be altered due to the flipped resistances. In this context, Eve remains unaware of when the switching operation transpires because both $b_A/b_B=1/0$ and $b_A/b_B=0/1$ produce an intermediate noise level, preventing Eve from accurately decoding the bits in these cases. This causes Eve to make decoding errors both when $b_A$ and $b_B$ are identical and when they are different.

Alice's decision tree mechanism is presented in Fig. \ref{fig:dectree}, where Bob also has the same one. The mechanism operates such that, by knowing the current state and the random bit, Alice has a priori knowledge about the expected voltage variance on the wire. If $L$ is used, which is the case when $S_A$ is $\mathscr{N}$ and $b_A=0$, or when $S_A=\mathscr{F}$ and $b_A=1$), the estimated variance is compared to the threshold $\gamma_1$. If the estimated variance is smaller, it is concluded that Bob used $L$; otherwise, Bob used $H$. Similarly, if $H$ is used (when $S_A=\mathscr{N}$ and $b_A=1$, or when $S_A=\mathscr{F}$ and $b_A=0$), the estimated variance is compared to the threshold $\gamma_2$. If the estimated variance is smaller, it is deduced that Bob used $L$; otherwise, Bob used $H$. Finally, knowing the resistance Bob used and $S_B$, Alice can determine Bob's bit, as both Alice and Bob change their states synchronously.

$S_A$ flips when $b_A/d_B=1/0$; similarly, $S_B$ flips when $d_A/b_B=1/0$, where $d_A$ and $d_B$ are Alice's detected bit by Bob and Bob's detected bit by Alice, respectively. In cases where Alice or Bob make an erroneous detection and mistakenly interpret $b_A/b_B=1/0$, the one who made the incorrect detection remains in their current state, while the other's state changes. This discrepancy between their states arises due to the mismatch caused by the erroneous detection. This state disparity introduces the potential for error propagation, where a single-bit error may lead to an average of two subsequent bit errors. The reason for this is that their states only realign when their random bits are $0/0$ or $1/1$ ($0.5$ probability) under specific conditions, as illustrated in Table \ref{tabletransition}. These mismatch cases and their probabilities are carefully examined in the following subsection.
\begin{table}[t!]
\centering
\caption{\label{tabletransition} Self-Correction of the State Mismatches}
\begin{tblr}{
  vline{1,2,3,4,5,6,7}={1.1pt},
  hline{1,2,6} = {1.1pt},
  hline{3-5} = {1-7}{},
  colspec = {X[c,m]X[c,m]X[c,m]X[c,m]X[c,m]X[c,m]X[c,m]},
  stretch = 0.3,
  rowsep = 3pt,
}
\SetCell{bg=lightgray}{\textbf{Present} \\ \textbf{State}\\\textbf{$S_A/S_B$}} & 
\SetCell{bg=lightgray}{\textbf{Random} \\ \textbf{Bits}\\\textbf{$b_A/b_B$}} & 
\SetCell{bg=lightgray}{\textbf{Resistors} \\ \resizebox{31.5pt}{!}{\textbf{Alice/Bob}}} & 
\SetCell{bg=lightgray}{\textbf{$b_A/d_B$}} & 
\SetCell{bg=lightgray}{\textbf{$d_A/b_B$}} &
\SetCell{bg=lightgray}{\textbf{Next} \\ \textbf{State}\\\textbf{$S_A/S_B$}} \\

$\mathscr{N}/\mathscr{F}$ & $0/0$ & $L/H$ & $0/1$ & $1/0$ & $\mathscr{N}/\mathscr{N}$ \\
$\mathscr{N}/\mathscr{F}$ & $1/1$ & $H/L$ & $1/0$ & $0/1$ & $\mathscr{F}/\mathscr{F}$ \\
$\mathscr{F}/\mathscr{N}$ & $1/1$ & $L/H$ & $1/0$ & $0/1$ & $\mathscr{N}/\mathscr{N}$ \\
$\mathscr{F}/\mathscr{N}$ & $0/0$ & $H/L$ & $0/1$ & $1/0$ & $\mathscr{F}/\mathscr{F}$ \\
\end{tblr}
\end{table}

\subsection{Probabilities of Mismatched States}
We define the probability of mismatches between $S_A$ and $S_B$ as follows. There are four possible mismatch scenarios: a transition to a mismatched state, either from $\mathscr{N}/\mathscr{N}$ or $\mathscr{F}/\mathscr{F}$, to $\mathscr{N}/\mathscr{F}$ or $\mathscr{F}/\mathscr{N}$. The probability of mismatch, $P_{mm}$, can be calculated as
\begingroup
\begin{align}
P_{mm}=2\big(P&(S_A/S_B = \mathscr{N}/\mathscr{N} \rightarrow \mathscr{N}/\mathscr{F})\quad&&\nonumber\\
+P&(S_A/S_B = \mathscr{F}/\mathscr{F} \rightarrow \mathscr{N}/\mathscr{F})\qquad\nonumber\\
+P&(S_A/S_B = \mathscr{N}/\mathscr{N} \rightarrow \mathscr{F}/\mathscr{N})\qquad\nonumber\\
+P&(S_A/S_B = \mathscr{F}/\mathscr{F} \rightarrow \mathscr{F}/\mathscr{N})\big), \label{Eq:7}\raisetag{12pt}
\end{align}
\endgroup
These terms are subsequently multiplied by two due to the probability of error propagation, which continues until the bits reach $b_A/b_B=0/0$ or $b_A/b_B=1/1$ as stated in Table \ref{tabletransition}. Considering that the bits are equally probable, and the likelihood of encountering $b_A/b_B=0/0 \text{ or } 1/1$ is $0.5\times0.5+0.5\times0.5=0.5$, it implies an average duration of two bits for self-correction to occur.

Table \ref{bigtable} provides insight into all potential outcomes for transmitted and detected bits under varying state conditions, along with explanations for their validity. For instance, in the first row of the table, where $S_A/S_B = \mathscr{N}/\mathscr{N} \rightarrow \mathscr{N}/\mathscr{F}$, it signifies that both Alice and Bob's previous state is $\mathscr{N}$. When $ b_A/b_B=0/0 $ and $ S_A/S_B=\mathscr{N}/\mathscr{N} $, we would typically anticipate $ \hat{\sigma}^2 < \gamma_1 $. However, if both Alice and Bob estimate $ \hat{\sigma}^2 > \gamma_1 $, Alice will decide $ d_B=1 $ and Bob will decide $ d_A=1 $. In this scenario, while $ b_A/d_B=0/1 $ and $ S_A $ remains unchanged, $ b_B/d_A=1/0 $ causes $ S_B $ to flip. Decisions highlighted in red indicate invalid outcomes, whereas those not highlighted are deemed valid. 
The reasoning behind their validity or invalidity is clarified in the last column. For example, decisions such as $d_B/d_A=0/0$ or $d_B/d_A=1/0$ are considered invalid because $S_B$ cannot flip if $d_A=0$, as $S_B$ only flips when $d_A/b_B=1/0$. Similarly, $d_B/d_A=0/1$ is not a feasible detection where $\hat{\sigma}^2$ does not remain consistent for both Alice and Bob, as $\hat{\sigma}^2<\gamma_1$ for Alice and $\hat{\sigma}^2>\gamma_1$ for Bob cannot occur simultaneously.

In state transition cases, only two possible detections exist. For instance, in the first state transition case in the table, when $b_A/b_B=0/0$ and $d_A/d_B=1/1$, it is considered a valid detection because $\hat{\sigma}^2>\gamma_1$ for both Alice and Bob. The corresponding probabilities in the table are defined as $p_{1A}=p_{1B}=P(\hat{\sigma}^2>\gamma_1)$ in this case. Two possible detections exist for all four state transition cases. These detection cases will be further defined and calculated in the following subsections.

We note that different previous states are not accounted for, as we focus on determining the probability of transitioning to the case of mismatch while the states are the same on both sides. Below, we sequentially define and compute these four probabilities.
\begin{table}[hbt!]
\centering
\caption{Random Bits and the Possible Decisions Made by Alice and Bob During Inconsistent State Transitions}
\label{bigtable}
\resizebox{1\linewidth}{!}{
\begin{tblr}{
  cell{1}{1} = {r=2}{},
  cell{1}{2} = {c=2}{},
  cell{1}{4} = {c=2}{},
  cell{1}{6} = {r=2}{},
  cell{3}{1} = {r=10}{},
  cell{3}{2} = {r=4}{},
  cell{3}{3} = {r=4}{},
  cell{7}{4} = {c=2}{},
  cell{8}{2} = {r=4}{},
  cell{8}{3} = {r=4}{},
  cell{12}{4} = {c=2}{},
  cell{13}{1} = {r=10}{},
  cell{13}{4} = {c=2}{},
  cell{14}{4} = {c=2}{},
  cell{15}{2} = {r=4}{},
  cell{15}{3} = {r=4}{},
  cell{19}{2} = {r=4}{},
  cell{19}{3} = {r=4}{},
  cell{23}{1} = {r=10}{},
  cell{23}{4} = {c=2}{},
  cell{24}{4} = {c=2}{},
  cell{25}{2} = {r=4}{},
  cell{25}{3} = {r=4}{},
  cell{29}{2} = {r=4}{},
  cell{29}{3} = {r=4}{},
  cell{33}{1} = {r=10}{},
  cell{33}{2} = {r=4}{},
  cell{33}{3} = {r=4}{},
  cell{37}{4} = {c=2}{},
  cell{38}{2} = {r=4}{},
  cell{38}{3} = {r=4}{},
  cell{42}{4} = {c=2}{},
  vline{3,5,7},
  vline{1,2,4,6,7}={1pt},
  hline{1,3,13,23,33,43} = {1-6}{1pt},
  hline{3-43} = {6-7}{},
  hline{2,7-8,12,14-15,19,24-25,29,37-38,42} = {2-5}{},
  hline{4-6,9-11,16-18,20-22,26-28,30-32,34-36,39-41} = {4-5}{},
  colspec = {X[c,m,35.42pt]X[c,m,7.8pt]X[c,m,7.8pt]X[c,m,7.8pt]X[c,m,7.8pt]X[m,106pt]},
  stretch = 0.6,
  rowsep = 3.45pt,
}
\SetCell{bg=lightgray}\textbf{State Transition} \resizebox{!}{\height}{\hspace{-3pt}$S^\text{prev} \rightarrow S$} & \SetCell{bg=lightgray}\textbf{Bits} &  & \SetCell{bg=lightgray}\textbf{Decision} & & \SetCell{bg=lightgray}\hspace{13pt}\textbf{Valid (\textcolor{ForestGreen}{\ding{51}}) or Invalid (\textcolor{red}{\ding{56}})} \\ 
 & \SetCell{bg=lightgray}$b_A$ & \SetCell{bg=lightgray}$b_B$ & \SetCell{bg=lightgray}$d_B$ & \SetCell{bg=lightgray}$d_A$\\
\SetCell{bg=lightgray}{
\parbox[c][0cm][c]{2cm}{\centering
\hspace{-21pt}$S_A/S_B$\\[2pt]
\hspace{-21pt}$=$\\[2pt]
\hspace{-21pt}$\mathscr{N}/\mathscr{N}$\\[2pt]
\hspace{-21pt}$\downarrow$\\[2pt]
\hspace{-21pt}$\mathscr{N}/\mathscr{F}$\\[6pt]
\resizebox{!}{10.5pt}{\hspace{-26pt}
$\begin{pmatrix}
S_A\text{ holds}\\
S_B\text{ changes}
\end{pmatrix}$}
}
} & 0 & 0 & \SetCell{bg=lightred}0 & \SetCell{bg=lightred}0 & $\hspace{-5pt}$\textcolor{red}{\textcolor{red}{\ding{56}}}$P(d_A=0|S_B\text{ changes})=0$\\
 &  &  & \SetCell{bg=lightred}0 & \SetCell{bg=lightred}1 & $\hspace{-5pt}$\textcolor{red}{\textcolor{red}{\ding{56}}}\resizebox{!}{5.7pt}{$P(\hat{\sigma}^2<\gamma_1)\cap P(\hat{\sigma}^2>\gamma_1)=0$}\\
 &  &  & \SetCell{bg=lightred}1 & \SetCell{bg=lightred}0 & $\hspace{-5pt}$\textcolor{red}{\textcolor{red}{\ding{56}}}$P(d_A=0|S_B\text{ changes})=0$\\
 &  &  & 1 & 1 & $\hspace{-5pt}$\textcolor{ForestGreen}{\ding{51}}$p_{1A}\cap p_{1B}>0$\\
 & \SetCell{bg=lightred}0 & \SetCell{bg=lightred}1 & \SetCell{bg=lightred}- & &$\hspace{-5pt}$\textcolor{red}{\textcolor{red}{\ding{56}}}$P(b_B=1|S_B\text{ changes})=0$ \\
 & 1 & 0 & \SetCell{bg=lightred}0 & \SetCell{bg=lightred}0 & $\hspace{-5pt}$\textcolor{red}{\textcolor{red}{\ding{56}}}$P(d_B=0|S_A\text{ holds})=0$\\
 &  &  & \SetCell{bg=lightred}0 & \SetCell{bg=lightred}1 & $\hspace{-5pt}$\textcolor{red}{\textcolor{red}{\ding{56}}}$P(d_B=0|S_A\text{ holds})=0$\\
 &  &  & \SetCell{bg=lightred}1 & \SetCell{bg=lightred}0 & $\hspace{-5pt}$\textcolor{red}{\textcolor{red}{\ding{56}}}$P(d_A=0|S_B\text{ changes})=0$\\
 &  &  & 1 & 1 & $\hspace{-5pt}$\textcolor{ForestGreen}{\ding{51}}$p_{2A}\cap p_{2B}>0$\\
 & \SetCell{bg=lightred}1 & \SetCell{bg=lightred}1 & \SetCell{bg=lightred}- & &$\hspace{-5pt}$\textcolor{red}{\ding{56}}$ P(b_B=1|S_B\text{ changes})=0$ \\
\SetCell{bg=lightgray}{
\parbox[c][0cm][c]{2cm}{\centering
\hspace{-21pt}$S_A/S_B$\\[2pt]
\hspace{-21pt}$=$\\[2pt]
\hspace{-21pt}$\mathscr{N}/\mathscr{N}$\\[2pt]
\hspace{-21pt}$\downarrow$\\[2pt]
\hspace{-21pt}$\mathscr{F}/\mathscr{N}$\\[6pt]
\resizebox{!}{10.5pt}{\hspace{-26pt}
$\begin{pmatrix}
S_A\text{ changes}\\
S_B\text{ holds}
\end{pmatrix}$}
}
} & \SetCell{bg=lightred}0 & \SetCell{bg=lightred}0 & \SetCell{bg=lightred}- & & $\hspace{-5pt}$\textcolor{red}{\ding{56}}$ P(b_A=0|S_A\text{ changes})=0$\\
 & \SetCell{bg=lightred}0 & \SetCell{bg=lightred}1 & \SetCell{bg=lightred}- & & $\hspace{-5pt}$\textcolor{red}{\ding{56}}$ P(b_A=0|S_A\text{ changes})=0$ \\
 & 1 & 0 & 0 & 0 & $\hspace{-5pt}$\textcolor{ForestGreen}{\ding{51}}$p_{5A}\cap p_{5B}>0$\\
 &  &  & \SetCell{bg=lightred}0 & \SetCell{bg=lightred}1 & $\hspace{-5pt}$\textcolor{red}{\ding{56}}$P(d_A=1|S_B\text{ holds})=0$\\
 &  &  & \SetCell{bg=lightred}1 & \SetCell{bg=lightred}0 & $\hspace{-5pt}$\textcolor{red}{\ding{56}}$P(d_B=1|S_A\text{ changes})=0$\\
 &  &  & \SetCell{bg=lightred}1 & \SetCell{bg=lightred}1 & $\hspace{-5pt}$\textcolor{red}{\ding{56}}$P(d_B=1|S_A\text{ changes})=0$\\
 & 1 & 1 & 0 & 0 & $\hspace{-5pt}$\textcolor{ForestGreen}{\ding{51}}$p_{6A}\cap p_{6B}>0$\\
 &  &  & \SetCell{bg=lightred}0 & \SetCell{bg=lightred}1 & $\hspace{-5pt}$\textcolor{red}{\ding{56}}\resizebox{!}{5.62pt}{$P(\hat{\sigma}^2<\gamma_2)\cap P(\hat{\sigma}^2>\gamma_2)=0$}\\
 &  &  & \SetCell{bg=lightred}1 & \SetCell{bg=lightred}0 & $\hspace{-5pt}$\textcolor{red}{\ding{56}}$P(d_B=1|S_A\text{ changes})=0$\\
 &  &  & \SetCell{bg=lightred}1 & \SetCell{bg=lightred}1 & $\hspace{-5pt}$\textcolor{red}{\ding{56}}$P(d_B=1|S_A\text{ changes})=0$\\
\SetCell{bg=lightgray}{
\parbox[c][0cm][c]{2cm}{\centering
\hspace{-21pt}$S_A/S_B$\\[2pt]
\hspace{-21pt}$=$\\[2pt]
\hspace{-21pt}$\mathscr{F}/\mathscr{F}$\\[2pt]
\hspace{-21pt}$\downarrow$\\[2pt]
\hspace{-21pt}$\mathscr{N}/\mathscr{F}$\\[6pt]
\resizebox{!}{10.5pt}{\hspace{-26pt}
$\begin{pmatrix}
S_A\text{ changes}\\
S_B\text{ holds}
\end{pmatrix}$}
}
} & \SetCell{bg=lightred}0 & \SetCell{bg=lightred}0 & \SetCell{bg=lightred}- & & $\hspace{-5pt}$\textcolor{red}{\ding{56}}$P(b_A=0|S_A\text{ changes})=0$ \\
 & \SetCell{bg=lightred}0 & \SetCell{bg=lightred}1 & \SetCell{bg=lightred}- & &$\hspace{-5pt}$\textcolor{red}{\ding{56}}$P(b_A=0|S_A\text{ changes})=0$ \\
 & 1 & 0 & 0 & 0 &$\hspace{-5pt}$\textcolor{ForestGreen}{\ding{51}}$p_{3A}\cap p_{3B}>0$\\
 &  &  & \SetCell{bg=lightred}0 & \SetCell{bg=lightred}1 &$\hspace{-5pt}$\textcolor{red}{\ding{56}}$P(d_A=1|S_B\text{ holds})=0$\\
 &  &  & \SetCell{bg=lightred}1 & \SetCell{bg=lightred}0 &$\hspace{-5pt}$\textcolor{red}{\ding{56}}$P(d_B=1|S_A\text{ changes})=0$\\
 &  &  & \SetCell{bg=lightred}1 & \SetCell{bg=lightred}1 &$\hspace{-5pt}$\textcolor{red}{\ding{56}}$P(d_B=1|S_A\text{ changes})=0$\\
 & 1 & 1 & 0 & 0 &$\hspace{-5pt}$\textcolor{ForestGreen}{\ding{51}}$p_{4A}\cap p_{4B}>0$\\
 &  &  & \SetCell{bg=lightred}0 & \SetCell{bg=lightred}1 &$\hspace{-5pt}$\textcolor{red}{\textcolor{red}{\ding{56}}}\resizebox{!}{5.62pt}{$P(\hat{\sigma}^2>\gamma_1)\cap P(\hat{\sigma}^2<\gamma_1)=0$}\\
 &  &  & \SetCell{bg=lightred}1 & \SetCell{bg=lightred}0 &$\hspace{-5pt}$\textcolor{red}{\ding{56}}$P(d_B=1|S_A\text{ changes})=0$\\
 &  &  & \SetCell{bg=lightred}1 & \SetCell{bg=lightred}1 &$\hspace{-5pt}$\textcolor{red}{\ding{56}}$P(d_B=1|S_A\text{ changes})=0$\\
\SetCell{bg=lightgray}{
\parbox[c][0cm][c]{2cm}{\centering
\hspace{-21pt}$S_A/S_B$\\[2pt]
\hspace{-21pt}$=$\\[2pt]
\hspace{-21pt}$\mathscr{F}/\mathscr{F}$\\[2pt]
\hspace{-21pt}$\downarrow$\\[2pt]
\hspace{-21pt}$\mathscr{F}/\mathscr{N}$\\[6pt]
\resizebox{!}{10.5pt}{\hspace{-26pt}
$\begin{pmatrix}
S_A\text{ holds}\\
S_B\text{ changes}
\end{pmatrix}$}
}
} & 0 & 0 & \SetCell{bg=lightred}0 & \SetCell{bg=lightred}0 &$\hspace{-5pt}$\textcolor{red}{\ding{56}}$P(d_A=0|S_B\text{ changes})=0$\\
 &  &  & \SetCell{bg=lightred}0 & \SetCell{bg=lightred}1 &$\hspace{-5pt}$\textcolor{red}{\textcolor{red}{\ding{56}}}\resizebox{!}{5.7pt}{$P(\hat{\sigma}^2>\gamma_2)\cap P(\hat{\sigma}^2<\gamma_2)=0$}\\
 &  &  & \SetCell{bg=lightred}1 & \SetCell{bg=lightred}0 &$\hspace{-5pt}$\textcolor{red}{\ding{56}}$P(d_A=0|S_B\text{ changes})=0$\\
 &  &  & 1 & 1&$\hspace{-5pt}$\textcolor{ForestGreen}{\ding{51}}$p_{7A}\cap p_{7B}>0$\\
 & \SetCell{bg=lightred}0 & \SetCell{bg=lightred}1 & \SetCell{bg=lightred}- & &$\hspace{-5pt}$\textcolor{red}{\ding{56}}$P(b_B=1|S_B\text{ changes})=0$  \\
 & 1 & 0 & \SetCell{bg=lightred}0 & \SetCell{bg=lightred}0 &$\hspace{-5pt}$\textcolor{red}{\ding{56}}$P(d_A=0|S_B\text{ changes})=0$\\
 &  &  & \SetCell{bg=lightred}0 & \SetCell{bg=lightred}1 &$\hspace{-5pt}$\textcolor{red}{\ding{56}}$P(d_B=0|S_A\text{ holds})=0$\\
 &  &  & \SetCell{bg=lightred}1 & \SetCell{bg=lightred}0 &$\hspace{-5pt}$\textcolor{red}{\ding{56}}$P(d_A=0|S_B\text{ changes})=0$\\
 &  &  & 1 & 1&$\hspace{-5pt}$\textcolor{ForestGreen}{\ding{51}}$p_{8A}\cap p_{8B}>0$\\
 & \SetCell{bg=lightred}1 & \SetCell{bg=lightred}1 & \SetCell{bg=lightred}- & &$\hspace{-5pt}$\textcolor{red}{\textcolor{red}{\ding{56}}}$P(b_B=1|S_B\text{ changes})=0$  
\end{tblr}}
\end{table}
\subsubsection{Calculation of $P(S_A/S_B = \mathscr{N}/\mathscr{N} \rightarrow \mathscr{N}/\mathscr{F})$}

\begin{figure*}[t!]
\centering
\medskip
\setlength{\fboxsep}{-2pt}
\fbox{\subfigure{
\includegraphics[trim={3pt 7pt 3pt 2pt},clip,width=0.233\linewidth]{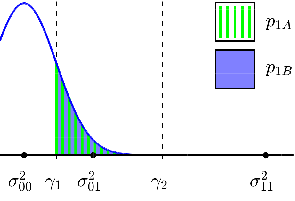} 
}}
\setlength{\fboxsep}{-2pt}
\fbox{\subfigure{
\includegraphics[trim={3pt 7pt 3pt 2pt},clip,width=0.233\linewidth]{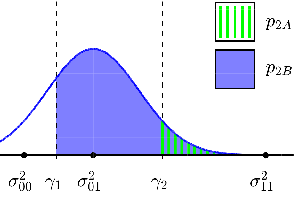}
}}
\setlength{\fboxsep}{-2pt}
\fbox{\subfigure{
\includegraphics[trim={3pt 7pt 3pt 2pt},clip,width=0.233\linewidth]{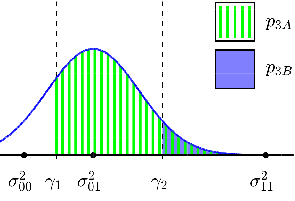} 
}}
\setlength{\fboxsep}{-2pt}
\fbox{\subfigure{
\includegraphics[trim={3pt 7pt 3pt 2pt},clip,width=0.233\linewidth]{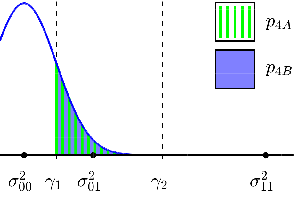}
}}
\\ \vspace{8pt}
\setlength{\fboxsep}{-2pt}
\fbox{\subfigure{
\includegraphics[trim={3pt 7pt 3pt 2pt},clip,width=0.233\linewidth]{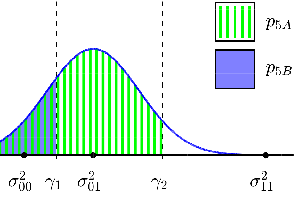} 
}}
\setlength{\fboxsep}{-2pt}
\fbox{\subfigure{
\includegraphics[,trim={3pt 7pt 3pt 2pt},clip,width=0.233\linewidth]{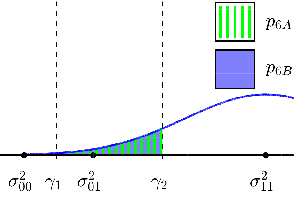}
}}
\setlength{\fboxsep}{-2pt}
\fbox{\subfigure{
\includegraphics[trim={3pt 7pt 3pt 2pt},clip,width=0.233\linewidth]{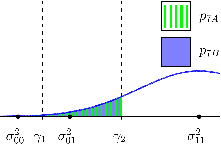} 
}}
\setlength{\fboxsep}{-2pt}
\fbox{\subfigure{
\includegraphics[trim={3pt 7pt 3pt 2pt},clip,width=0.233\linewidth]{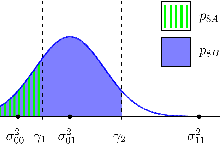}
}}
\caption{Demonstration of $p_{i}=p_{iA}\cap p_{iB}$. }
\label{Fig:subprobs}
\end{figure*}

We define this probability as the likelihood of a transition from $S_A/S_B=\mathscr{N}/\mathscr{N}$ to $\mathscr{N}/\mathscr{F}$, which can be calculated as follows:

\begin{align}
\begin{aligned}
    P(S_A/S_B=\mathscr{N}/\mathscr{N}\rightarrow \mathscr{N}/\mathscr{F})=\qquad\qquad\qquad\qquad\qquad\qquad\\[5pt]
    P(d_B/d_A=1/1\;|\;b_A/b_B=0/0\;\&\;S^\text{prev}_A/S^\text{prev}_B=\mathscr{N}/\mathscr{N})&\\[5pt]
    \times P(b_A/b_B=0/0)P(S^\text{prev}_A/S^\text{prev}_B=\mathscr{N}/\mathscr{N})&\\[5pt]
    + P(d_B/d_A=1/1\;|\;b_A/b_B=1/0\;\&\;S^\text{prev}_A/S^\text{prev}_B=\mathscr{N}/\mathscr{N})&\\[5pt]
    \times P(b_A/b_B=1/0)P(S^\text{prev}_A/S^\text{prev}_B=\mathscr{N}/\mathscr{N})&,\label{Eq:8}
\end{aligned}
\end{align}
where $P(b_A/b_B=0/0)=P(b_A/b_B=1/0)=0.25$. Noting that $S^\text{prev}_A$ and $S^\text{prev}_B$ should be the same and can be either $\mathscr{N}$ or $\mathscr{F}$, $P(S^\text{prev}_A/S^\text{prev}_B=\mathscr{N}/\mathscr{N})=0.5$. Substituting these probability values in (\ref{Eq:8}) and calling the others $p_1$ and $p_2$, respectively, we obtain
\begin{align}\label{Eq:5v2}
P(S_A/S_B=\mathscr{N}/\mathscr{N}\rightarrow \mathscr{N}/\mathscr{F})=0.125(p_1+p_2).
\end{align}
$p_1$ and $p_2$ are calculated by further defining $p_{iA}$ and $p_{iB}$ as the probabilities that Alice and Bob make the decision stated in $p_{i}$, respectively, where $i\in\{1,2,\dots,8\}$ and $p_{i}=p_{iA}\cap p_{iB}$. These probabilities are shown in Fig. \ref{Fig:subprobs}. The probabilities $p_{1A}$ and $p_{1B}$ can be obtained to find $p_1$ as
\begin{align}
\begin{aligned}
    p_{1A}&=P(d_B=1\;|\;b_A/b_B=0/0\;\&\;S^\text{prev}_A/S^\text{prev}_B=\mathscr{N}/\mathscr{N})\\[5pt]
    &=P(\hat{\sigma}^2 > \gamma_1),\label{Eq:99}\raisetag{10pt}
\end{aligned}
\end{align}\vspace{-10pt}
\begin{align}
\begin{aligned}
    p_{1B}&=P(d_A=1\;|\;b_A/b_B=0/0\;\&\;S^\text{prev}_A/S^\text{prev}_B=\mathscr{N}/\mathscr{N})\\
    &=P(\hat{\sigma}^2 > \gamma_1).\label{Eq:1010}\raisetag{10pt}
\end{aligned}
\end{align}
In this scenario, when $b_A/b_B=0/0$ and $S^\text{prev}_A/S^\text{prev}_B=\mathscr{N}/\mathscr{N}$, which corresponds (\ref{Eq:99}), it follows that $\sigma^2=\sigma_{00}^2$ and
$d_B$ may equal $1$ solely if Alice estimates $\hat{\sigma}^2 > \gamma_1$. Similarly, in (\ref{Eq:1010}), $d_A$ can equal $1$ only if Bob estimates $\hat{\sigma}^2 > \gamma_1$. Since the intersection of these possibilities is again $\hat{\sigma}^2 > \gamma_1$, $p_1$ can be calculated as
\begingroup
\setlength{\abovedisplayskip}{10pt}
\setlength{\belowdisplayskip}{10pt}
\begin{equation}\label{Eq:110}
\begin{aligned}
    p_1=p_{1A}\cap p_{1B}=P(\hat{\sigma}^2 > \gamma_1) = Q\left(\dfrac{\gamma_1-\sigma_{00}^2}{\sqrt{2\sigma_{00}^4/N}}\right).
\end{aligned}
\end{equation}
\endgroup
On the other hand, $p_{2B}$ corresponds to a larger measurement interval than $p_{2A}$. We define these two probabilities as
\begingroup
\setlength{\abovedisplayskip}{14pt}
\setlength{\belowdisplayskip}{14pt}
\begin{align}
\begin{aligned}
    p_{2A}&=P(d_B=1\;|\;b_A/b_B=1/0\;\&\;S^\text{prev}_A/S^\text{prev}_B=\mathscr{N}/\mathscr{N}),\\[5pt]
    &=P(\hat{\sigma}^2 > \gamma_2),\\[5pt]
    p_{2B}&=P(d_A=1\;|\;b_A/b_B=1/0\;\&\;S^\text{prev}_A/S^\text{prev}_B=\mathscr{N}/\mathscr{N}).\\[5pt]
    &=P(\hat{\sigma}^2 > \gamma_1).
\end{aligned}\raisetag{10pt}
\end{align}
\endgroup
Here, $\sigma^2=\sigma_{01}^2$ since $b_A/b_B=1/0$, and the value of $d_B$ may equal $1$ only if Alice estimates $\hat{\sigma}^2 > \gamma_2$. Similarly, $d_A$ can equal $1$ solely if Bob estimates $\hat{\sigma}^2 > \gamma_1$. Given that these scenarios intersect exclusively within the interval $\hat{\sigma}^2 > \gamma_2$ where $\gamma_2>\gamma_1$, we proceed to calculate $p_2$ as follows:
\begin{equation}
\setlength\abovedisplayskip{10pt}
\vspace{5pt}
\begin{aligned}\label{Eq:112}
    p_2=p_{2A}\cap p_{2B}=P(\hat{\sigma}^2 > \gamma_2) = Q\left(\dfrac{\gamma_2-\sigma_{01}^2}{\sqrt{2\sigma_{01}^4/N}}\right).
\end{aligned}
\end{equation}
\subsubsection{Calculation of $P(S_A/S_B = \mathscr{F}/\mathscr{F} \rightarrow \mathscr{N}/\mathscr{F})$}
The probability of a transition from $S_A/S_B=\mathscr{F}/\mathscr{F}$ to $\mathscr{N}/\mathscr{F}$ can be calculated as follows:
\begin{align}
\begin{aligned}
    P(S_A/S_B=\mathscr{F}/\mathscr{F}\rightarrow \mathscr{N}/\mathscr{F})=\qquad\qquad\qquad\qquad\qquad\qquad\\[5pt]
    P(d_B/d_A=0/0\;|\;b_A/b_B=1/0\;\&\;S^\text{prev}_A/S^\text{prev}_B=\mathscr{F}/\mathscr{F})&\\[5pt]
    \times P(b_A/b_B=1/0)P(S^\text{prev}_A/S^\text{prev}_B=\mathscr{F}/\mathscr{F})&\\[5pt]
    + P(d_B/d_A=0/0\;|\;b_A/b_B=1/1\;\&\;S^\text{prev}_A/S^\text{prev}_B=\mathscr{F}/\mathscr{F})&\\[5pt]
    \times P(b_A/b_B=1/1)P(S^\text{prev}_A/S^\text{prev}_B=\mathscr{F}/\mathscr{F})&,\label{Eq:10}
\end{aligned}
\end{align}
where $P(b_A/b_B=1/1)=0.25$ and $P(S^\text{prev}_A/S^\text{prev}_B=\mathscr{F}/\mathscr{F})=0.5$. Thus, (\ref{Eq:10}) becomes as follows:
\begin{equation}\label{Eq:11}
\begin{aligned}
P(S_A/S_B=\mathscr{F}/\mathscr{F}\rightarrow \mathscr{N}/\mathscr{F})=0.125(p_3+p_4),
\end{aligned}
\end{equation}
Similar approach in (\ref{Eq:110}) and (\ref{Eq:112}) is used to calculate $p_3$ and $p_4$ as
\begin{equation}
\begin{aligned}\label{Eq:115}
    p_3&=p_{3A}\cap p_{3B}=P(\hat{\sigma}^2 > \gamma_2) = Q\left(\dfrac{\gamma_2-\sigma_{01}^2}{\sqrt{2\sigma_{01}^4/N}}\right),\\[2pt]
    p_4&=p_{4A}\cap p_{4B}=P(\hat{\sigma}^2 > \gamma_1) = Q\left(\dfrac{\gamma_1-\sigma_{00}^2}{\sqrt{2\sigma_{00}^4/N}}\right),
\end{aligned}
\end{equation}
where $p_{iA}$ and $p_{iB}$ values for $i\in\{3,\dots,8\}$ are given in the Appendix.
\subsubsection{Calculation of $P(S_A/S_B = \mathscr{N}/\mathscr{N} \rightarrow \mathscr{F}/\mathscr{N})$}
The likelihood of a transition from $S_A/S_B=\mathscr{N}/\mathscr{N}$ to $\mathscr{F}/\mathscr{N}$ can be calculated as follows:
\begingroup
\setlength{\abovedisplayskip}{10pt}
\setlength{\belowdisplayskip}{5pt}
\begin{align}
\begin{aligned}\label{Eq:14}
    P(S_A/S_B=\mathscr{N}/\mathscr{N}\rightarrow \mathscr{F}/\mathscr{N})=\qquad\qquad\qquad\qquad\qquad\quad\;\;\\[5pt]
    P(d_B/d_A=0/0\;|\;b_A/b_B=1/0\;\&\;S^\text{prev}_A/S^\text{prev}_B=\mathscr{N}/\mathscr{N})&\\[5pt]
    \times P(b_A/b_B=1/0)P(S^\text{prev}_A/S^\text{prev}_B=\mathscr{N}/\mathscr{N})&\\[5pt]
    + P(d_B/d_A=0/0\;|\;b_A/b_B=1/1\;\&\;S^\text{prev}_A/S^\text{prev}_B=\mathscr{N}/\mathscr{N})&\\[5pt]
    \times P(b_A/b_B=1/1)P(S^\text{prev}_A/S^\text{prev}_B=\mathscr{N}/\mathscr{N})&.
\end{aligned}
\end{align}
\endgroup
Similar approaches in the previous calculation of the conditional probability can be applied to further simplify (\ref{Eq:14}) as
\begin{equation}\label{Eq:15}
\begin{aligned}
P(S_A/S_B=\mathscr{N}/\mathscr{N}\rightarrow \mathscr{F}/\mathscr{N})=0.125(p_5+p_6),
\end{aligned}
\end{equation}
where,
\begin{equation}
\begin{aligned}\label{Eq:120}
    p_5&=p_{5A}\cap p_{5B}=P(\hat{\sigma}^2 < \gamma_1) = Q\left(\dfrac{\sigma_{01}^2-\gamma_1}{\sqrt{2\sigma_{01}^4/N}}\right),\\[5pt]
    p_6&=p_{6A}\cap p_{6B}=P(\hat{\sigma}^2 < \gamma_2) = Q\left(\dfrac{\sigma_{11}^2-\gamma_2}{\sqrt{2\sigma_{11}^4/N}}\right).
\end{aligned}
\end{equation}
\subsubsection{Calculation of $P(S_A/S_B = \mathscr{F}/\mathscr{F} \rightarrow \mathscr{F}/\mathscr{N})$}
The probability of a transition from $S_A/S_B=\mathscr{F}/\mathscr{F}$ to $\mathscr{F}/\mathscr{N}$ can be computed as follows:
\begin{align}\label{Eq:16}
\begin{aligned}
    P(S_A/S_B=\mathscr{F}/\mathscr{F}\rightarrow \mathscr{F}/\mathscr{N})=\qquad\qquad\qquad\qquad\qquad\qquad\\[5pt]
    P(d_B/d_A=1/1\;|\;b_A/b_B=0/0\;\&\;S^\text{prev}_A/S^\text{prev}_B=\mathscr{F}/\mathscr{F})& \\[5pt]
    \times P(b_A/b_B=0/0)P(S^\text{prev}_A/S^\text{prev}_B=\mathscr{F}/\mathscr{F})&\\[5pt]
    + P(d_B/d_A=1/1\;|\;b_A/b_B=1/0\;\&\;S^\text{prev}_A/S^\text{prev}_B=\mathscr{F}/\mathscr{F})& \\[5pt]
    \times P(b_A/b_B=1/0)P(S^\text{prev}_A/S^\text{prev}_B=\mathscr{F}/\mathscr{F})&.
\end{aligned}\raisetag{-5pt}
\end{align}
Considering similar simplifications in the previous case, (\ref{Eq:16}) becomes
\begin{equation}\label{Eq:17}
\begin{aligned}
P(S_A/S_B=\mathscr{F}/\mathscr{F}\rightarrow \mathscr{F}/\mathscr{N})=0.125(p_7+p_8),
\end{aligned}
\end{equation}
where,
\begin{equation}
\begin{aligned}\label{Eq:123}
    p_7&=p_{7A}\cap p_{7B}=P(\hat{\sigma}^2 < \gamma_2) = Q\left(\dfrac{\sigma_{11}^2-\gamma_2}{\sqrt{2\sigma_{11}^4/N}}\right),\\[5pt]
    p_8&=p_{8A}\cap p_{8B}=P(\hat{\sigma}^2 < \gamma_1) = Q\left(\dfrac{\sigma_{01}^2-\gamma_1}{\sqrt{2\sigma_{01}^4/N}}\right).
\end{aligned}
\end{equation}
To summarize, all the $p_i$ values are represented as follows:
\begin{equation}\label{Eq:25}
\begin{aligned}
    p_1=p_4=P(\hat{\sigma}^2 > \gamma_1) = Q\left(\dfrac{\gamma_1-\sigma_{00}^2}{\sqrt{2\sigma_{00}^4/N}}\right),\\[6pt]  %
    p_2=p_3=P(\hat{\sigma}^2 > \gamma_2) = Q\left(\dfrac{\gamma_2-\sigma_{01}^2}{\sqrt{2\sigma_{01}^4/N}}\right),\\[6pt]
    p_5=p_8=P(\hat{\sigma}^2 < \gamma_1) = Q\left(\dfrac{\sigma_{01}^2-\gamma_1}{\sqrt{2\sigma_{01}^4/N}}\right),\\[6pt]
    p_6=p_7=P(\hat{\sigma}^2 < \gamma_2) = Q\left(\dfrac{\sigma_{11}^2-\gamma_2}{\sqrt{2\sigma_{11}^4/N}}\right).\\[2pt]
\end{aligned}
\end{equation}
Eventually, we derive the overall mismatch probability ($P_{mm}$) by substituting (\ref{Eq:5v2}), (\ref{Eq:11}), (\ref{Eq:15}) and (\ref{Eq:17}) in (\ref{Eq:7}) as
\begin{align}\label{Eq:24}
\begin{aligned}
    P_{mm}=&0.25(p_1+p_2+p_3+p_4+p_5+p_6+p_7+p_8)\\[5pt]
    =&0.5(p_1+p_2+p_5+p_6)\\[5pt]
    =&0.5\Bigg[Q\left(\dfrac{\gamma_1-\sigma_{00}^2}{\sqrt{2\sigma_{00}^4/N}}\right)
    +Q\left(\dfrac{\gamma_2-\sigma_{01}^2}{\sqrt{2\sigma_{01}^4/N}}\right)\quad\\
    &\;\ +Q\left(\dfrac{\sigma_{01}^2-\gamma_1}{\sqrt{2\sigma_{01}^4/N}}\right)
    +Q\left(\dfrac{\sigma_{11}^2-\gamma_2}{\sqrt{2\sigma_{11}^4/N}}\right)\Bigg].\\[2pt]
\end{aligned}
\end{align}
Let us simplify this term by normalizing all these terms with respect to the smallest one, $\sigma_{00}^2$. Recalling that $\gamma_1=\beta\sigma_{00}^2$, $\gamma_2=\kappa\sigma_{00}^2$, $\sigma_{01}^2=(2\alpha/(1+\alpha))\sigma_{00}^2$, and $\sigma_{11}^2=\alpha\sigma_{00}^2$, (\ref{Eq:24}) becomes
\begingroup
\setlength{\abovedisplayskip}{10pt}
\setlength{\belowdisplayskip}{10pt}
\begin{align}\label{Eq:255}
\begin{aligned}
    P_{mm}=0.5\Bigg[Q\Bigg(\dfrac{\beta-1}{\sqrt{2/N}}\Bigg)
    +Q\Bigg(\dfrac{\kappa-\big(\frac{2\alpha}{1+\alpha}\big)}{\big(\frac{2\alpha}{1+\alpha}\big)\sqrt{2/N}}\Bigg)&&\\[3pt]
    +Q\Bigg(\dfrac{\big(\frac{2\alpha}{1+\alpha}\big)-\beta}{\big(\frac{2\alpha}{1+\alpha}\big)\sqrt{2/N}}\Bigg)
    +Q\Bigg(\dfrac{\alpha-\kappa}{\alpha\sqrt{2/N}}\Bigg)\Bigg].&&
\end{aligned}
\end{align}
\endgroup

\subsection{Theoretical BEP of Flip-KLJN Scheme}
We define the total BEP of the Flip-KLJN scheme as
\begin{align}
    P_b=P_{mm} P_{b^{mm}} + P_m P_{b^m},
\end{align}
where $P_{b^{mm}}$, $P_m$, and $P_{b^m}$ hold for the BEP in mismatch, probability of a match, and the BEP in the match. Here, we note that $P_{b^{mm}} = 1$ because Alice and Bob will make wrong decisions if their states do not match and $P_m = 1 - P_{mm}$ with the rule of sum. Moreover, $P_{b^m}$ is already explained in detail and calculated in \cite{9980386} as
\begingroup
\setlength{\abovedisplayskip}{10pt}
\setlength{\belowdisplayskip}{10pt}
\begin{align}\label{Eq:28}
\begin{aligned}
    P_{b^m}=0.25\Bigg[Q\Bigg(\dfrac{\beta-1}{\sqrt{2/N}}\Bigg)
    +Q\Bigg(\dfrac{\kappa-\big(\frac{2\alpha}{1+\alpha}\big)}{\big(\frac{2\alpha}{1+\alpha}\big)\sqrt{2/N}}\Bigg)&&\\[5pt]
    +Q\Bigg(\dfrac{\big(\frac{2\alpha}{1+\alpha}\big)-\beta}{\big(\frac{2\alpha}{1+\alpha}\big)\sqrt{2/N}}\Bigg)
    +Q\Bigg(\dfrac{\alpha-\kappa}{\alpha\sqrt{2/N}}\Bigg)\Bigg].&&
\end{aligned}
\end{align}
\endgroup
As seen from (\ref{Eq:255}) and (\ref{Eq:28}), $P_{mm}=2P_{b^m}$. In the light of these, the overall BEP is obtained as follows:
\begin{equation}\label{Eq:29}
    \begin{aligned}
        P_b=3P_{b^m}-2({P_{b^m}})^2.
    \end{aligned}
\end{equation}

\vspace{-10pt}



\section{Numerical Results}
In this section, we present our extensive computer simulations, comparing the BER performances of our proposed Flip-KLJN, Flip-KLJN with JVCD, classical KLJN, and classical KLJN with JVCD schemes. We assume $\alpha=10$, unless stated otherwise. The values for $\beta$, $\kappa$, $\eta$, and $\xi$ that optimize the BER are determined through computer simulations.

\begin{figure}[t!]
    \centering
    \includegraphics[width=\columnwidth]{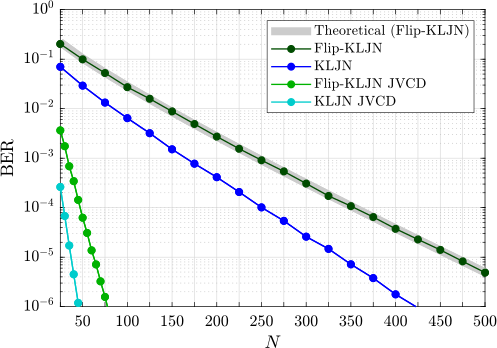}
    \caption{BER performances for different KLJN schemes and varying $N$.} 
    \label{fig:changeN}
\end{figure}

In Fig. \ref{fig:changeN}, we present the BER performance for both the KLJN and Flip-KLJN schemes. Both demonstrates a reduction in BER with an increase in the number of samples $N$, indicating enhanced performance with larger sample sizes at either Alice's or Bob's end. However, a slight disparity exists between these designs due to error propagation, as stated earlier. It is noteworthy that error propagation typically ceases after an average of four bits upon the initial mismatch between states. To address this performance gap, the JVCD \cite{9980386,4578679} is employed to mitigate bit errors and reduce occurrences of mismatch. The Flip-KLJN JVCD and KLJN JVCD schemes notably exhibit substantially lower BERs compared to other schemes that does not have this enhanced detector, particularly at higher $N$ values, signifying superior error reduction performance. The results underscore the advantages of employing the JVCD. Moreover, to achieve the same BER for Flip-KLJN JVCD as KLJN JVCD, we need to increase $N$ by a factor of approximately $1.7$. For example, KLJN JVCD with $N=25$ achieves nearly the same BER as Flip-KLJN JVCD with $N=40$, so $N$ should be increased by a factor of $40/25=1.6$. Similarly, KLJN JVCD with $N=45$ and Flip-KLJN JVCD with $N=75$ show a similar BER, requiring an increase factor of $75/45 \approx 1.7$. This increment in $N$ will slightly reduce the effective key rate, but since the Flip-KLJN scheme already doubles the key rate, the proposed design still achieves a higher key rate overall. Additionally, after $N=75$, BER falls below $10^{-6}$, which is sufficiently low for most applications \cite{saez2013errors}, making further increases in $N$ unnecessary. Furthermore, it is evident from Fig. \ref{fig:changeN} that our theoretical calculations and simulation results are consistent.

\begin{figure}[t]
    \centering
    \includegraphics[width=\columnwidth]{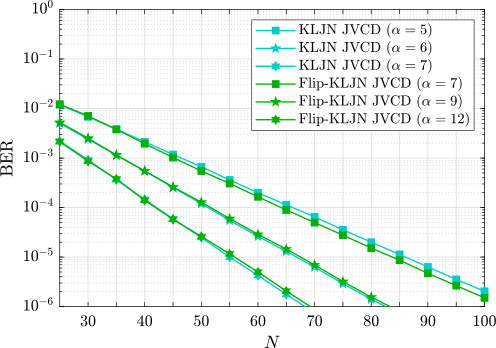}
    \caption{BER performances for different KLJN schemes and varying $N$.}
    \label{fig:changealphaN}
\end{figure}

Increasing $\alpha$ may be another solution to address the performance gap between Flip-KLJN JVCD and KLJN JVCD schemes. The computer simulation results are presented in Fig. \ref{fig:changealphaN} for both designs with JVCD using different $\alpha$ values. One can observe that nearly the same BER can be achieved if $\alpha$ is increased by $2$, $3$, and $5$ for the Flip-KLJN JVCD scheme when the KLJN JVCD scheme has $\alpha$ values of $5$, $6$, and $7$, respectively. Notably, this increment in $\alpha$ becomes more significant when the KLJN JVCD scheme uses larger values of $\alpha$. However, it is also worth noting that increasing $\alpha$ to match the exact BER of the KLJN JVCD scheme may not be necessary if the BER drops below $10^{-6}$. This low BER performance can easily be achieved with $\alpha>10$ and $N>100$, which are generally practical values for real-life implementations \cite{saez2013errors,4578679}.

\begin{figure}[t]
    \centering
    \includegraphics[width=\columnwidth]{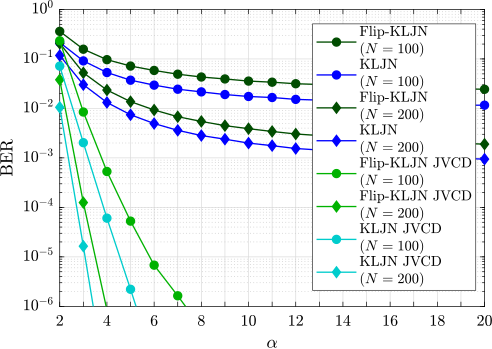}
    \caption{BER performances for different KLJN schemes and varying $\alpha$.}
    \label{fig:changealpha}
\end{figure}
In Fig. \ref{fig:changealpha}, we highlight the impact of the parameter $\alpha$ on the BER for both schemes at $N=100$ and $N=200$. As $\alpha$ increases, the BER decreases for all schemes, indicating improved performance. On the other hand, increasing $\alpha$ introduces several parasitic effects and affect system stability. Excessive $\alpha$ amplifies high-frequency components, making the system more vulnerable to parasitic capacitance and inductance, which distort signals and degrade performance. If artificial noise generators are used, higher $\alpha$ increases signal energy, requiring more power. Moreover, for both Flip-KLJN and KLJN schemes, increasing the number of samples from $100$ to $200$ significantly reduces the BER. The schemes utilizing the JVCD detector exhibit notably lower BERs. Additionally, we observe that the performance gap between KLJN JVCD and Flip-KLJN JVCD schemes diminishes as $N$ increases. Error floors are also examined for designs without JVCD schemes due to statistical decision errors, and these error floors decrease to values below $10^{-10}$ with the help of JVCD \cite{4578679}.

\begin{figure}[t]
    \centering
    \includegraphics[width=\columnwidth]{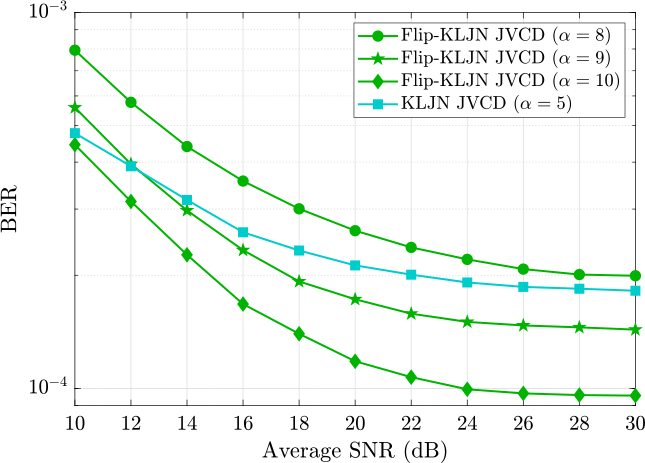}
    \caption{The BER performance of the KLJN JVCD and Flip-KLJN JVCD schemes under various average SNR values:}
    \label{fig:noiseBERKLJN}
\end{figure}
     
The BER performance of the Flip-KLJN JVCD and KLJN JVCD schemes under varying average SNR conditions is presented in Fig. \ref{fig:noiseBERKLJN}. In this analysis, the average signal power is defined as the mean of the three possible noise variance levels in the KLJN loop. The calculation uses practical physical parameters: Boltzmann constant \(k = 1.38 \times 10^{-23}\,\text{J/K}\), temperature \(T = 300\,\text{K}\), bandwidth \(B = 10^6\,\text{Hz}\), and \(R_L = 1000\,\Omega\). As previously mentioned, the JVCD detector utilizes both voltage and current measurements, and measurement noise is added to both signal types during sampling. The BER decreases with increasing SNR, indicating improved signal detection in the presence of reduced measurement noise. While the Flip-KLJN JVCD scheme initially exhibits a higher BER than KLJN JVCD scheme due to potential error propagation introduced by the flipping operation, this drawback can be mitigated by increasing the resistance ratio \(\alpha\). As shown in Fig. \ref{fig:noiseBERKLJN}, larger values of \(\alpha\) significantly improve BER performance by enhancing the distinguishability between different noise levels, thereby improving robustness against measurement noise.

\begin{figure}[t]
    \centering
    \includegraphics[width=\columnwidth]{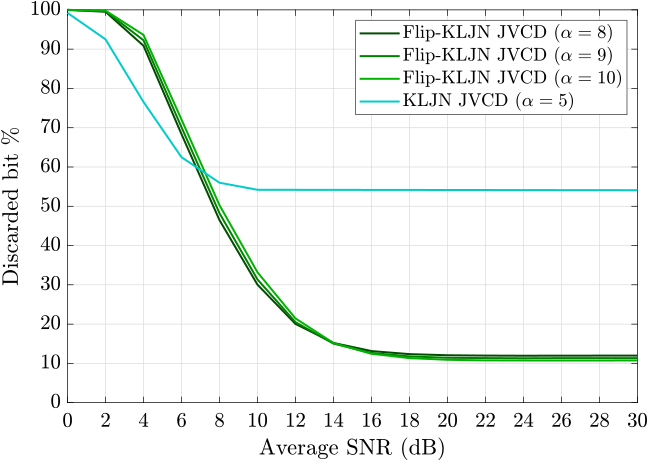}
    \caption{The percentage of the discarded bits of the KLJN JVCD and Flip-KLJN JVCD schemes under various average SNR values:}
    \label{fig:sucBitRat}\vspace{6pt}
\end{figure}

The percentage of discarded bits for the Flip-KLJN JVCD and KLJN JVCD schemes is illustrated in Fig. \ref{fig:sucBitRat}. In the KLJN JVCD scheme, a substantial portion of bits are discarded either due to insecure noise power combinations or mismatches between voltage and current estimations. In contrast, the Flip-KLJN JVCD scheme significantly reduces the number of discarded bits, as all noise power levels are utilized securely. As the SNR increases, the percentage of discarded bits in the Flip-KLJN JVCD scheme decreases rapidly, stabilizing below $10\%$ beyond $16$ dB. This demonstrates the Flip-KLJN JVCD scheme’s capability to maintain a higher effective key rate while preserving the same level of security.

It is also evident from Fig. \ref{fig:sucBitRat} that the discarded bit percentage in the KLJN JVCD design reaches its minimum value at around $10$ dB, whereas the Flip-KLJN JVCD design reaches its minimum at approximately $16$ dB. Although the Flip-KLJN JVCD scheme requires a slightly higher SNR to reach optimal performance, this does not present a significant drawback. For instance, at an average SNR of $10$ dB, the Flip-KLJN JVCD scheme discards approximately $30\%$ of the bits, compared to about $55\%$ in the KLJN JVCD design. Given that the Flip-KLJN JVCD scheme effectively doubles the key rate, it still achieves superior overall performance. However, when the average SNR falls below $10$ dB, both schemes become ineffective due to the high rate of discarded bits.

\vspace{-4pt}
\section{Conclusion}

In conclusion, this paper has introduced the innovative Flip-KLJN design as a strategic solution to the limitations of conventional KLJN schemes in using insecure bits. The proposed approach confounds potential eavesdroppers by systematically inverting resistances during key generation, achieving unconditional security resembling quantum secrecy. The primary contributions of this work include the development of the Flip-KLJN design, its integration with various detectors to mitigate BER discrepancies, and an increased key rate. Our results demonstrated that the key rate can be expanded by a factor of two compared to the classical KLJN. Looking ahead, future research may explore the generalization of communication schemes with multiple resistors, establish information-theoretical bounds on data rate and secrecy capacity, address practical challenges such as wire resistance and sampling imperfections, generate key books for added unconditional security, employment of colored noise, and development of coding schemes to enhance error performance. Most importantly, the proposed scheme may enable two-way secure data transfer, where both entities simultaneously send and receive bits by using randomization algorithms for their information bits in the future.
\vspace{-7pt}
\section*{Acknowledgments}
We would like to express our deepest gratitude to Professor Laszlo B. Kish, whose invaluable guidance, support, and expertise were instrumental in the completion of this work.

{\appendix[Calculation of $p_{iA}$ and $p_{iB}$ for All Values of $i$]
$p_{3A}$ and $p_{3B}$ in (\ref{Eq:115}) can be calculated as
\begin{equation}
\begin{aligned}
    p_{3A}&=P(d_B=0\;|\;b_A/b_B=1/0\;\&\;S^\text{prev}_A/S^\text{prev}_B=\mathscr{F}/\mathscr{F})\\
    &=P(\hat{\sigma}^2 > \gamma_1),\\
    p_{3B}&=P(d_A=0\;|\;b_A/b_B=1/0\;\&\;S^\text{prev}_A/S^\text{prev}_B=\mathscr{F}/\mathscr{F})\\
    &=P(\hat{\sigma}^2 > \gamma_2).
\end{aligned}
\end{equation}
Moreover, $p_{4A}$ and $p_{4B}$ in (\ref{Eq:115}) is computed as
\begin{equation}
\begin{aligned}
    p_{4A}&=P(d_B=0\;|\;b_A/b_B=1/1\;\&\;S^\text{prev}_A/S^\text{prev}_B=\mathscr{F}/\mathscr{F})\\
    &=P(\hat{\sigma}^2 > \gamma_1),\\
    p_{4B}&=P(d_A=0\;|\;b_A/b_B=1/1\;\&\;S^\text{prev}_A/S^\text{prev}_B=\mathscr{F}/\mathscr{F})\\
    &=P(\hat{\sigma}^2 > \gamma_1).
\end{aligned}
\end{equation}
$p_{5A}$ and $p_{5B}$ in (\ref{Eq:120}) can be defined as
\begin{equation}
\begin{aligned}
    p_{5A}&=P(d_B=0\;|\;b_A/b_B=1/0\;\&\;S^\text{prev}_A/S^\text{prev}_B=\mathscr{N}/\mathscr{N})\\
    &=P(\hat{\sigma}^2 < \gamma_2),\\
    p_{5B}&=P(d_A=0\;|\;b_A/b_B=1/0\;\&\;S^\text{prev}_A/S^\text{prev}_B=\mathscr{N}/\mathscr{N})\\
    &=P(\hat{\sigma}^2 < \gamma_1).
\end{aligned}
\end{equation}
$p_{6A}$ and $p_{6B}$ in (\ref{Eq:120}) is calculated as
\begin{equation}
\begin{aligned}
    p_{6A}&=P(d_B=0\;|\;b_A/b_B=1/1\;\&\;S^\text{prev}_A/S^\text{prev}_B=\mathscr{N}/\mathscr{N})\\
    &=P(\hat{\sigma}^2 < \gamma_2),\\
    p_{6B}&=P(d_A=0\;|\;b_A/b_B=1/1\;\&\;S^\text{prev}_A/S^\text{prev}_B=\mathscr{N}/\mathscr{N})\\
    &=P(\hat{\sigma}^2 < \gamma_2).
\end{aligned}
\end{equation}
$p_{7A}$ and $p_{7B}$ in (\ref{Eq:123}) can be computed as
\begin{equation}
\begin{aligned}
    p_{7A}&=P(d_B=1\;|\;b_A/b_B=0/0\;\&\;S^\text{prev}_A/S^\text{prev}_B=\mathscr{F}/\mathscr{F})\\
    &=P(\hat{\sigma}^2 < \gamma_2),\\
    p_{7B}&=P(d_A=1\;|\;b_A/b_B=0/0\;\&\;S^\text{prev}_A/S^\text{prev}_B=\mathscr{F}/\mathscr{F})\\
    &=P(\hat{\sigma}^2 < \gamma_2).
\end{aligned}
\end{equation}
Lastly, $p_{8A}$ and $p_{8B}$ in (\ref{Eq:123}) is computed as
\begin{equation}
\begin{aligned}
    p_{8A}&=P(d_B=1\;|\;b_A/b_B=1/0\;\&\;S^\text{prev}_A/S^\text{prev}_B=\mathscr{F}/\mathscr{F})\\
    &=P(\hat{\sigma}^2 < \gamma_1),\\
    p_{8B}&=P(d_A=1\;|\;b_A/b_B=1/0\;\&\;S^\text{prev}_A/S^\text{prev}_B=\mathscr{F}/\mathscr{F})\\
    &=P(\hat{\sigma}^2 < \gamma_2).
\end{aligned}
\end{equation}
}
\vspace{-10pt}

\ifCLASSOPTIONcaptionsoff
  \newpage
\fi

\bibliographystyle{IEEEtran}
\bibliography{ref}

\begin{IEEEbiography}[{\includegraphics[width=1in,height=1.25in,clip,keepaspectratio]{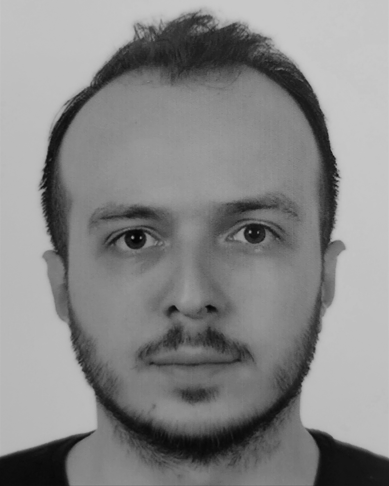}}]{\textbf{Recep A. Tasci}} (Graduate Student Member, IEEE) received the B.S. degree (with High Honors) in Electrical and Electronics Engineering from Istanbul Medipol University and the M.S. degree (with High Honors) in Electrical and Electronics Engineering from Koç University, Turkey, in 2020 and 2022, respectively. He is currently pursuing the Ph.D. degree at Koç University and is a Research Fellow at Medipol University, Turkey. His research interests include wireless communications, reconfigurable intelligent surfaces, channel modeling, signal processing, and zero-power and thermal noise communications. He has been serving as a reviewer for IEEE Transactions on Green Communications and Networking, IEEE Transactions on Wireless Communications, and IEEE Transactions on Vehicular Technology.
\end{IEEEbiography}

\begin{IEEEbiography}[{\includegraphics[width=1in,height=1.25in,clip,keepaspectratio]{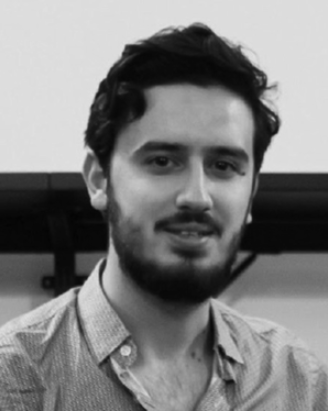}}]{\textbf{Ibrahim Yildirim}} (Member, IEEE) received the B.S. and M.S. degrees from Istanbul Technical University, Turkey, in 2017 and 2019, respectively. He received his Ph.D. degree from Koç University, Turkey. He is currently a Postdoctoral Research Fellow at the Broadband Communications Research Lab at McGill University, Canada. He served as a Research and Teaching Assistant at Istanbul Technical University from 2018 to 2023. He received the Exemplary Reviewer Award of the IEEE Transactions on Communications in 2021 and is a recipient of the Best Paper Award from the IEEE LATINCOM 2020. Additionally, he was awarded the Best Master Thesis Award by the IEEE Communications Society Turkey Section in 2021 and the Best Graduation Project Award by the Electrical Engineers Branch of Turkey in 2017. His current research interests include MIMO systems and reconfigurable intelligent surfaces. He has been serving as a Reviewer for IEEE Journal On Selected Areas In Communications, IEEE Transactions on Communications, IEEE Transactions on Vehicular Technology, and IEEE Communications Letters.
\end{IEEEbiography}

\begin{IEEEbiography}[{\includegraphics[width=1in,height=1.25in,clip,keepaspectratio]{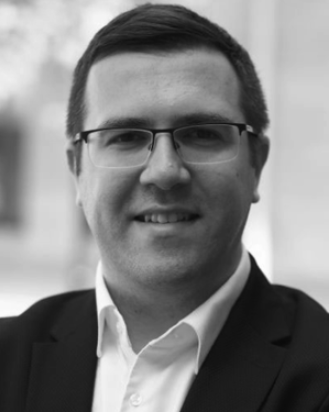}}]{\textbf{Ertugrul Basar}} (Fellow, IEEE) received the B.S. degree (with High Honors) from Istanbul University, Turkey, in 2007 and the M.Sc. and Ph.D. degrees from Istanbul Technical University, Turkey, in 2009 and 2013, respectively.
He is a Professor of Wireless Systems at the Department of Electrical Engineering, Tampere University, Finland. He was the founding director of the Communications Research and Innovation Laboratory (CoreLab) at Koç University, Istanbul, Turkey. Before joining Tampere University, he held positions at Koç University (2018-2025) and Istanbul Technical University (2009-2018). In the past, he had visiting positions at Princeton University, Princeton, NJ, USA, as a Visiting Research Collaborator (2011-2022) and at Ruhr University Bochum, Bochum, Germany, as a Mercator Fellow (2022).
Prof. Basar’s primary research interests include beyond 5G and 6G wireless systems, MIMO systems, index modulation, reconfigurable intelligent surfaces, waveform design, zero-power and thermal noise communications, software-defined radio, physical layer security, quantum key distribution systems, and signal processing/deep learning for communications. He is an inventor of around 15 pending/granted patents on future wireless technologies.
\end{IEEEbiography}

\vfill

\end{document}